\documentclass[seceq]{ptptex}

\usepackage{graphicx}
\usepackage{wrapft}

\title{
Induced Gravity in Deconstructed Space 
at Finite Temperature
}

\author{Nahomi \textsc{Kan}%
\footnote{E-mail: kan@yamaguchi-jc.ac.jp}
}
\inst{Yamaguchi Junior College, 1346-2 Daidou, Hofu-shi, Yamaguchi-ken
, Japan
}

\abst{
We study self-consistent cosmological solutions for an Einstein universe in a graph-based induced gravity model.
Especially, we demonstrate specific results for cycle graphs.
}

\begin{document}
\maketitle

\section{Pre-history}

\subsection{Induced Gravity}
The idea of induced Gravity is,
``Gravity emerges from the quantum effect of matter fields.''
The one-loop effective action can be expressed as the form: 
\begin{equation}
-\frac{1}{2}\int\frac{dt}{t}\sum_i {\rm Tr}\exp\left[-(-\nabla^2+M^2_i)t\right].
\end{equation}
In curved D-dimensional spacetime,
the trace part including the D-dimensional laplacian becomes
\begin{equation}
{\rm Tr}\exp\left[-(-\nabla^2)t\right]=\frac{\sqrt{g}}{(4\pi)^{D/2}}t^{-D/2}
(a_0+a_1 t+\cdots),
\end{equation}
where
coefficients depend on the background fields 
and $a_1$ 
leads to the Einstein-Hilbert term.
%

\subsection{Dimensional Deconstruction, Spectral Graph Theory and Our story thus far}
Dimensional Deconstruction (DD)\cite{Deconstruction} is equivalent to 
a higher-dimensional theory with discretized extra dimensions
at a low energy scale.
%
%
An $N$-sided polygon, called a ``moose" diagram, is used to describe this theory. 
%
Four-dimensional fields are assigned to vertices and edges of
this diagram. 
%
%
In $N \to \infty$,
DD leads to a five-dimensional theory,
where the extra space is a circle.  

In general, the moose diagram does not necessarily have a continuum limit 
and the diagram is susceptible of a complicated connection,
which is a {\it graph}. 
Therefore,
DD can be generalized to {\sl Field Theory on a Graph}.\cite{KSJMP}
We have constructed models of one-loop finite induced gravity by using several
graphs.\cite{KSPTP}
With the help of knowledge in {\sl Spectral Graph Theory},
we can easily find that the UV divergent terms concern the graph laplacian in DD. 
%

\section{Self-consistent Einstein Universe\cite{EU}} 
The metric of the static Einstein Universe is given by 
\begin{equation}
ds^2=-dt^2+a^2(d\chi^2+\sin^2\chi(d\theta^2+\sin^2\theta d\phi^2)),
\end{equation}
\begin{wrapfigure}{r}{5cm}
\includegraphics[height=5cm]{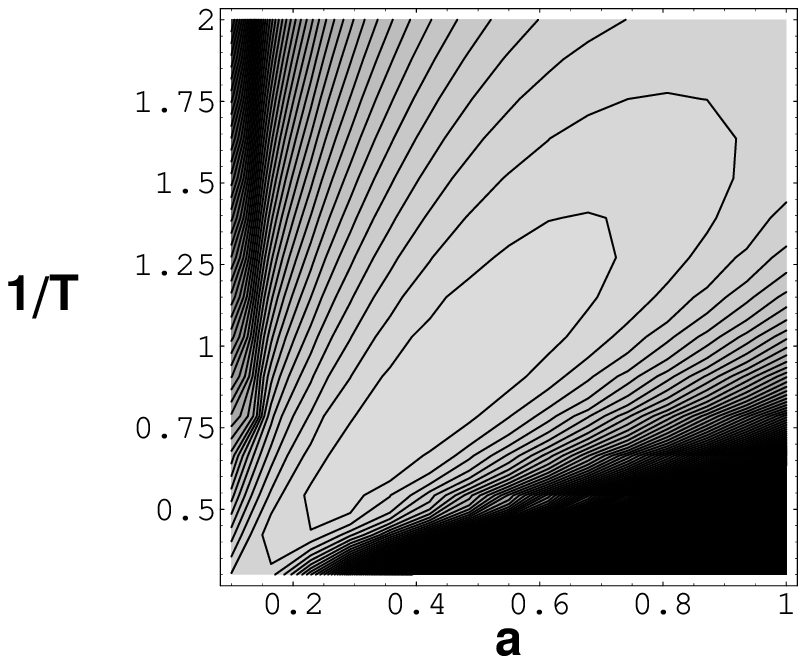}
\caption{%
A contour plot of $\beta F$ in the first model.
A solution of Einstein equation can be found at the maximum.
}
\label{fig1}
\smallskip
\smallskip
\smallskip
\smallskip
\includegraphics[height=5cm]{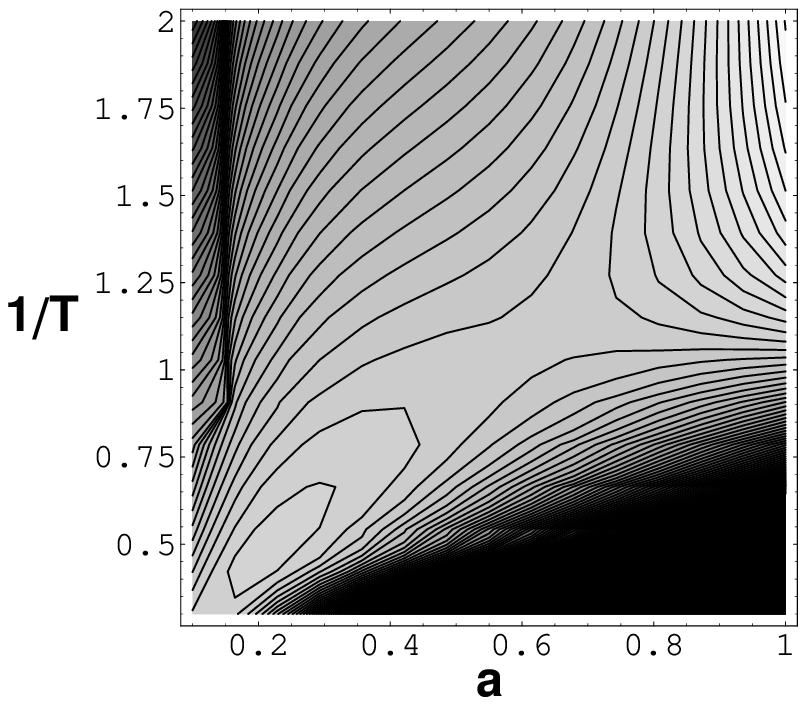}
\caption{%
A contour plot of $\beta F$ in the second model.
Two solutions of Einstein equation can be found at the maximum and at the saddle
point.
}
\label{fig2}
\end{wrapfigure}
where $a$ is the scale factor.
At finite temperature $T$, the one-loop effective action is regarded as free energy $F(a,\beta)$
and Einstein equation becomes  
\begin{equation}
\frac{\partial(\beta F)}{\partial \beta}=\frac{\partial(\beta F)}{\partial a}=0,
\label{Einstein-equation}
\end{equation}
where $\beta\equiv 1/T$.
In the first model, 
scalar fields are on 8 $C_{N/2}$, vector fields on 4 $C_N$ and Dirac fermions on 2 $C_{N/2}$ + 3 $C_N$.
%
In the second model, 
scalar fields are on 16 $C_{N/4}$ + 2 $C_{N/2}$, vector fields on 5 $C_N$ and Dirac
fermions on 4 $C_{N/4}$ + 3 $C_N$ + 2 $C_{N/2}$. Here $C_N$ denotes a cycle graph
with $N$ vertices, equivalent to an $N$-sided polygon. In each model,
Newton's constant and cosmological constant are calculable 
and are not given by hand. 
%
%

We exhibit $\beta F$
for the first model in Fig \ref{fig1}
and for the second in Fig \ref{fig2}, for large $N$.
The horizontal axis indicates the scale factor $a$ 
while the vertical one indicates the inverse of temperature $T$. 
The scale of each axis is in the unit of $N/f$.
In the first model, cosmological constant is zero 
and the solution 
can be found at the maximum of $\beta F$, corresponding to be
in Casimir regime.\cite{EU}
In the second model,
the solution in Casimir regime and the solution in Planck regime\cite{EU} are found.

\section{Summary} 
We have studied self-consistent Einstein Universe in the `graph theory space'.
The solution can be systematically obtained with the help of the graph structure.
%
%
%
%

\section*{Acknowledgements}
I would like to thank K.~Shiraishi for collaboration in this work,
and also the organizers of ICGA8.
%

\end{document}